# Influence of Matrix Composition on Microstructural Yielding and Vickers Hardness in Phase Separated Glasses

Nicholas L. Clark[1], Shih-Yi Chuang[1], John C. Mauro[1*]

[1]Department of Materials Science and Engineering, The Pennsylvania State University,

University Park, Pennsylvania 16802, USA

*Corresponding Author: jcm426@psu.edu

Abstract:

The relationship between matrix phase composition and microstructure yielding in phase separated calcium aluminosilicate glasses is investigated. Varying the treatment temperature of a phase separated glass results in glasses with different microstructures and matrix compositions. The impact of matrix composition on hardness depends on the mode of microstructural deformation. In glasses that deform via the droplet-densification mechanism, decreasing the matrix silica content results in a decrease in hardness due to the weakest-link effect and greater amounts of incongruent yielding of the phases. This relationship may act as an additional source of the indentation size effect in phase separated glasses. In glasses that deform via droplet-coalescence, decreasing the matrix silica content results in increasing hardness due to increased coalescence of the droplet phase during yielding.

1. Introduction

The relationship between phase separation and mechanical properties has been the subject of renewed discussion. Our recent work demonstrated basic relationships between microstructural yielding and the hardness of a phase separated glass for two archetypal microstructures.[1] Two distinct yielding mechanisms, droplet-coalescence and droplet-densification, were identified and found to occur in the acicular and spherical microstructures, respectively. However, a broad range of microstructures that span these two extremes may be synthesized within a single composition by varying the heat treatment temperature.[2] Assuming the treatment time is sufficient to establish a steady-state matrix composition, the composition of the droplet and matrix phases vary according to the lever rule at the given treatment temperature.[2–4] In the case of calcium aluminosilicates, lowering the treatment temperature increases the droplet fraction and decreases the silica content of the matrix.[2] The composition of the silica-rich droplets is relatively insensitive to the treatment temperature due to the strong left-skew of the immiscibility dome. No experimental study, to our knowledge, has directly related the properties of the matrix phase to hardness. However, matrix composition has been demonstrated to influence viscosity,[5] and is also predicted to influence crack propagation.[6,7] Therefore, the next step towards developing a relationship between phase separation and hardness in glasses is to understand the role phase composition plays during yielding.

The relationships between composition and hardness of homogeneous silicate glasses have already been thoroughly studied. These relationships arise from the formation of non-bridging oxygens (NBOs) with increasing modifier content,[8,9] changes in network connectivity,[10,11] and changes in the deformation mechanisms that are linked to the overall silica content.[10–12] Increasing the silica content generally increases hardness, while increasing the modifier-to-silica ratio tends to lower hardness. Given that preferential yielding of the soft matrix phase is responsible for the lower hardness of phase separated CAS glasses,[1] it is likely that altering the properties of the matrix phase will have some influence on the hardness.

2. Materials and Methods

Seven phase separated samples of a calcium aluminosilicate glass were synthesized by isothermally heat treating at different temperatures. This systematically varies the droplet fraction and by extension the matrix composition. The glass had a post-melt composition of 21.91CaO–6.37Al$_2$O$_3$–71.72SiO$_2$ (mol%) as measured via ICP–AES. All samples came from the same 300g melt. A constant treatment time of 30 minutes was used for each sample, as this period is long enough to ensure the microstructure has reached a steady-state composition.[2] After the prescribed treatment, the samples were air quenched to room temperature to avoid further nucleation and growth of the microstructure during cooling. This process generates additional thermal stress within the samples, so they were further annealed at 700°C for 30 minutes and allowed to cool slowly overnight. For ease of discussion, the samples will be referred to by their sample IDs rather than their specific treatment parameters. The sample IDs and corresponding treatment parameters are given in Table 1.

After heat treatment, the samples were mounted in epoxy disks and ground flat using 120 grit silicon carbide paper. The samples were then progressively polished down to 0.25 μm with diamond slurry. Vickers indentation was used to measure the indentation hardness of the glasses (Mitutoyo HM-200). A total of 20 indents at loads of 0.05, 0.1, 0.2, 0.3, 0.5, and 1.0 kgf were made in the sample. The indents were allowed to sit for 24 h before measuring their size using optical microscopy. All indentation testing was generally conducted in the same sitting to minimize variation in humidity and temperature. The indents were made at 22%–27% Relative Humidity (RH), and the RH varied from 20%–30% during the 24-h resting period.

The microstructures of each sample were imaged inside and outside the indent using Field Emission Scanning Electron Microscopy (FEI Verios 460L). The indented microstructure was imaged by tilting the sample by 22.5° during imaging so that one of the indent walls was perpendicular to the electron beam. At least five indents for each test force were imaged in this way when calculating microstructure statistics. From these images, the specific interfacial area (SIA), droplet fraction ($\phi$), and mean droplet size ($r$) were calculated.

For glasses with spherical droplets, the linear matrix strain ($\varepsilon_m$) was calculated using

$$\varepsilon_m = \frac{l_i - l_o}{l_i} \tag{1}$$

where $l_i$ and $l_o$ are the average inter-droplet distance inside and outside the indent, respectively. $l$ can be approximated from the previously measured microstructure statistics using

$$l = \sqrt{\frac{2}{3\phi}} r. \tag{2}$$

Likewise, the linear droplet strain ($\varepsilon_d$) was calculated using

$$\varepsilon_d = \frac{r_i - r_o}{r_i}. \tag{3}$$

From these values, the yield-congruency ratio ($\beta$) can be calculated as

$$\beta = \frac{\varepsilon_d}{\varepsilon_m}. \tag{4}$$

Further details regarding these kinds of measurements and calculations can be found in our previous works.[1,2] Additionally, the composition of the matrix phase was estimated at each temperature from the droplet fraction using a conservation of mass balance and assuming the droplet phase had a composition of 92– 6– 2 (SiO$_2$/CaO/Al$_2$O$_3$).[2,13]

## 3 Results

### 3.1 Matrix Composition

The droplet fraction of each base microstructure as a function of temperature is shown in Figure 1. The estimated matrix SiO$_2$, CaO, and Al$_2$O$_3$ contents for each glass are given in Table 2.

Using these compositions, an effective network former-to-modifier ratio, $R^*_{mat}$, can be calculated as

$$R^*_{mat} = \frac{[SiO_2]}{([CaO] - [Al_2O_3])} \qquad (5)$$

$R^*_{mat}$ is thus a measure of the amount of NBOs present in the matrix glass. Figure 2 shows the matrix silica content and $R^*_{mat}$ as a function of treatment temperature. $R^*_{mat}$ and the matrix silica content monotonically increase with treatment temperature. Therefore, increasing the treatment temperature increases the silica content and silica-to-modifier ratio within the matrix phase.

### 3.2 Hardness

Figure 3 shows the hardness at each load plotted against the treatment temperature. A non-monotonic relationship between hardness and temperature is observed for all loading conditions. Increasing the treatment temperature from 930°C to 1000°C results in a systematic decrease in hardness. However, increasing the treatment temperature beyond 1000°C results in an increase in hardness. The effect appears less pronounced as the applied load is increased; however, this is an artifact of the y-axis scaling. When the hardness values for each load are plotted individually the trend becomes more apparent. An example of this effect is shown for the 0.5kgf test force in Figure 4.

Increasing the treatment temperature has three fundamental effects on the microstructure. First, a higher treatment temperature increases the coarsening rate and leads to larger droplets at constant soak time. As demonstrated in our previous work, no systematic relationship between the initial mean droplet size and hardness exists in this system.[1] Second, increasing the treatment temperature decreases the droplet fraction. And third, increasing the treatment temperature results in increasingly spherical and symmetric droplet morphologies.

### 3.3 Droplet Strain

Figure 5 shows the linear droplet strain, $\varepsilon_d$, plotted as a function of applied load. Hallmarks of both droplet densification and droplet coalescence are apparent. In glasses T5, T6, and T7 a load-dependent decrease in droplet size occurs, suggesting they deform through droplet densification (Figure 6). In glasses T1, T2, and T3 droplet size increases with load relative to the base microstructure, suggesting they deform through droplet-coalescence (Figure 7). Droplet size in glass T4 initially displays both load-dependent increases and decreases but generally decreases relative to the base microstructure.

Figure 8 shows the linear matrix strain, $\varepsilon_m$, as a function of applied load for the DD glasses. Like $\varepsilon_d$, the magnitude of $\varepsilon_m$ increases with load except for glass T4, where $\varepsilon_m$ begins to increase beyond 0.3 kgf. This anomalous behaviour may be indicative of the onset of coalescence; however, this requires further investigation.

### 3.4 SIA

Figures 9 and 10 show the SIA and relative change in SIA as a function of applied load, respectively. Mirroring the droplet size changes in Figure 10, glasses T1-T3 display an initial decrease in SIA followed by a load-dependent increase. Glasses T5-T7 show an initial increase in SIA at low loads, and do not display a notable load dependency. Glass T4 does not show any systematic variation in SIA at any load.

### 3.5 Microstructure Yielding Mechanisms

Using the criteria developed in our previous work,[1] glasses T1, T2, and T3 deform through the droplet-coalescence mechanism, and glasses T4, T5, T6, and T7 deform through the droplet-densification mechanism. These groupings will be referred to as the droplet-coalescence glasses (DC glasses) and droplet-densification glasses (DD glasses). A small amount of coalescence occurs in glass T4 beyond 0.3kgf; however, the primary means of droplet deformation appears to be through the droplet-densification mechanism.

## 4 Discussion

### 4.1 Yield Congruency

Figure 11 shows the yield-congruency ratio, $\beta$, for the DD glasses versus the applied load.

Like the spherical microstructure seen in our previous work,[1] $\beta < 1$ which indicates preferential yielding of the matrix phase. A few important trends are apparent. First, β generally increases with the treatment temperature at low loads (Figure 12). This implies that increasing the modifier content of the matrix phase results in the matrix deforming more relative to the droplet phase. This is unsurprising, as increasing a glass's modifier content typically results in a decrease in hardness.

Second, $\beta$ initially increases at low load before plateauing around 0.3-0.5kgf for glasses T5, T6, and T7. $\beta$ for glass T4 does not plateau below 1.0kgf. $\beta$ is a measure of the relative contributions of the droplet and matrix deformations, therefore a plateau in $\beta$ indicates that the rate that the phases are deforming relative to each other is no longer changing with load.

In addition to composition, the mechanism of indentation-induced deformation in glasses is known to vary with the depth, or size, of the indent. Due to its relatively low activation barrier, densification is the primary means of plastic deformation at low loads.[12, 14, 15] In pure or high-silica glasses, densification occurs through redistribution of silica ring sizes and reduction in pore size.[16–19] Modified silicates possess significantly less free volume, and as a result densification occurs through $Q^n$ reorganization.[20–22] Ring redistribution occurs through changes in the inter-tetrahedral bond angles, while $Q^n$ reorganization involves breaking and re-forming Si-O bonds. Therefore, the activation barrier for densification is expected to increase with the introduction of modifiers.[12, 15] Therefore, densification of the silica-rich droplets is expected to have a lower activation barrier than densification in the modifier-rich matrix.

If this is the case, then why is $\beta$ smallest at low loads? The overall deformation would be expected to proceed from the lowest to the highest activation barrier process. Therefore, densification of the droplet phase should happen first, followed by matrix densification. This would result in the opposite of the trend observed, with $\beta$ being highest at low loads then decreasing with load.

This anomalous behavior is likely due to the unique requirements for deformation in a heterogeneous material. Consider a circumstance where loading *only* resulted in densification within the droplet phase. This results in a decrease in droplet size and by extension a decrease in droplet fraction. Decreasing the droplet volume fraction inherently requires the matrix volume fraction to increase. This would require the inflow of the matrix glass, presumably through shear flow. Therefore, the activation barrier for densification of the droplet phase is irrelevant to the overall deformation because the process is limited by the higher activation barrier for matrix shear flow.

This provides additional context as to why, despite phase separation in this system simply being a conservative reorganization of the bonding environment, the overall deformation in the DD glasses is controlled by the "weakest link". Because the densification within the droplets is restricted by the matrix phase, and the droplets are too dilute to coalesce, they have no way to dissipate energy through plastic deformation. Therefore, the silica that is taken from the matrix to form the droplet phase is effectively "wasted", and results in decreased hardness.

### 4.1 Effect of Matrix Silica Content on Hardness

A normalized hardness can be calculated by dividing the hardness at each load by the estimated silica content of the matrix from Table 2. The hardness-load curves before and after normalization are shown in Figures 13 and 14, respectively.

Normalizing the hardness to silica content results in glasses that deformed through droplet-densification collapsing onto a single curve, while glasses that deformed through coalescence diverge from the DD glass curve. The hardness and normalized hardness-load curves for the DD glasses are shown in Figures 15 and 16, respectively.

The hardness-load curves exhibit the indentation size effect up to 1kgf. In homogeneous glasses, hardness typically plateaus after a certain load. Previous work in the CAS system showed that loads beyond 0.1kgf did not exhibit the ISE.[12] The ISE is typically attributed to friction,[23] environmental effects,[24,25] cracking and subsurface damage,[26] or a change in the deformation mechanism.[27–29] However, the actual contribution of friction is believed to be relatively negligible in microindentations. Humidity and temperature did not appreciably change between testing different samples; therefore, the contribution of differing environmental effects is likely negligible. Additionally, no evidence of subsurface damage was observed in any sample at any load. Therefore, the persistent ISE beyond 0.1kgf observed in Figure 16 may be attributable to changes in the deformation mechanism.

Previously, $\beta$ was shown to vary with load at low loads before eventually plateauing at higher loads. $\beta$ is a measure of the relative contributions of droplet and matrix deformations. A constant $\beta$ indicates that the rate that the relative deformations of the phases is not varying with load. Therefore until $\beta$ plateaus, the deformation mechanism is changing and likely results in an additional ISE. This is demonstrated when the silica-normalized hardness is further normalized by $\beta$ (Figure 17).

Normalized hardness for T5, T6, and T6 shows no notable load-dependence after their load-$\beta$ curves plateau at 0.2kgf. However, glass T4, which did not display a plateau in $\beta$, continues to show the ISE after normalization. This is likely due to the onset of coalescence beyond 0.3kgf. Load-dependent normalized hardness is still observed below 0.2kgf, though. This may indicate that the of other sources of the ISE that tend to dominate at low loads are still active.

### 4.3 Effect of Matrix Silica Content on Coalescence

The silica content also affects the extent of coalescence. Figure 18 shows the hardness of the coalescence glasses as a function of change in SIA. As expected, hardness generally increases with the reduction in SIA within each glass. However, when the silica-normalized hardness is plotted against the reduction in SIA (Figure 19), the hardness-SIA curve of each individual glass collapses into a single curve.

This implies that the droplet phase's ability to undergo coalescence is inversely related to the silica content of the matrix phase. In order for two droplets to merge, the matrix glass between them must flow out of the space between them. Likewise, for a high aspect ratio droplet to deform into a lower aspect ratio shape, the matrix glass must cooperatively deform with the droplet. Therefore, the droplets' ability to participate in coalescence is inversely proportional to the strength of the matrix phase. The weaker the matrix phase, the more it will yield during indentation. The more the matrix phase yields, the more opportunity the droplet phase must deform through coalescence. A similar phenomenon was noted as the source of the characteristic incongruent yielding within the DD mechanism.

### 5 Conclusions

The results of this work demonstrate the influence of matrix phase composition on microstructural yielding and the hardness of phase separated glasses. High-temperature treatment results in the formation of dilute, spherical, silica-rich droplets at the expense of the modifier-enriched matrix phase. Indentation of these glasses results in incongruent yielding of the droplet and matrix phases. The silica contained within the droplet phase does not contribute to the overall yielding that occurs during indentation, resulting in an overall hardness that is dependent on the properties of the matrix. As the treatment temperature is lowered, the droplet fraction increases and the matrix silica content decreases, which leads to lower hardness.

As the treatment temperature continues to be lowered the microstructure transitions from the more dilute, spherical, droplet microstructure to a higher density microstructure with more acicular-shaped droplets. The change in microstructure allows the silica that was taken out of the matrix phase to dissipate energy during yielding via the droplet-coalescence mechanism. Now, a stronger matrix phase is detrimental to the overall hardness because deformation of the droplet phase is ultimately restricted by the ability of the matrix to deform with the droplets. Therefore, as the treatment temperature is reduced further, the hardness begins to increase because the matrix becomes easier to deform. Load-dependent yield congruency is likely responsible for the persistence of the ISE at higher loads in phase separated glasses. While not the only source of the ISE, fluctuations in the yield-congruency ratio should be considered as a possible source of load-dependent hardness when glasses are suspected or known to be phase separated.

6    Acknowledgements

This material is based upon work supported by the National Science Foundation under Grant No. 1762275. The authors are grateful to Dr. Jane Cook and Dr. Seong Kim for valuable discussions, and to Julie Andrews and Wes Auker for assistance with data collection.

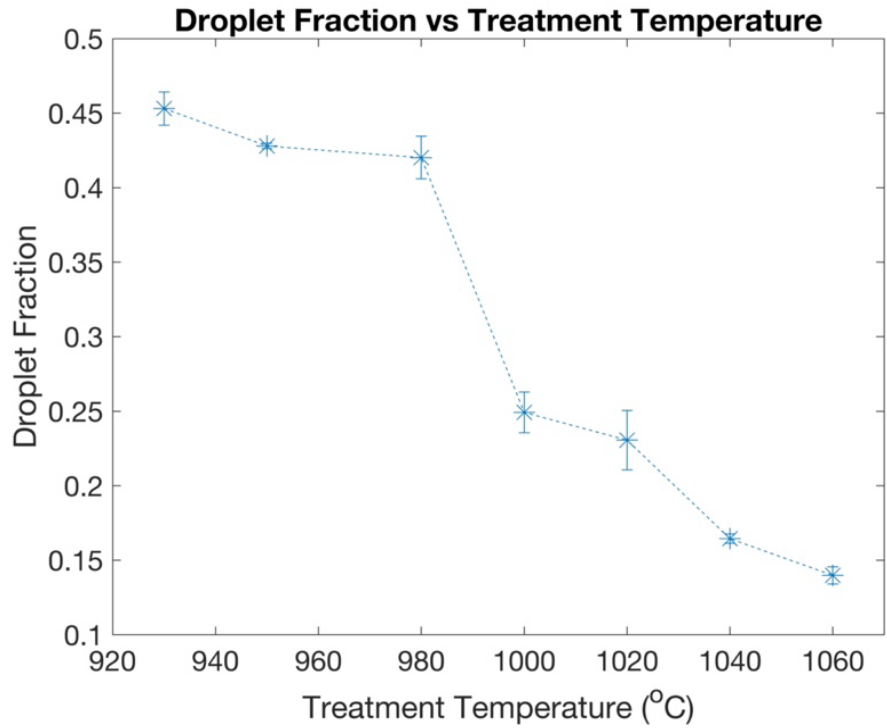

Figure 1: Droplet fraction vs treatment temperature for the T-series glasses. Droplet fraction monotonically decreases with treatment temperature

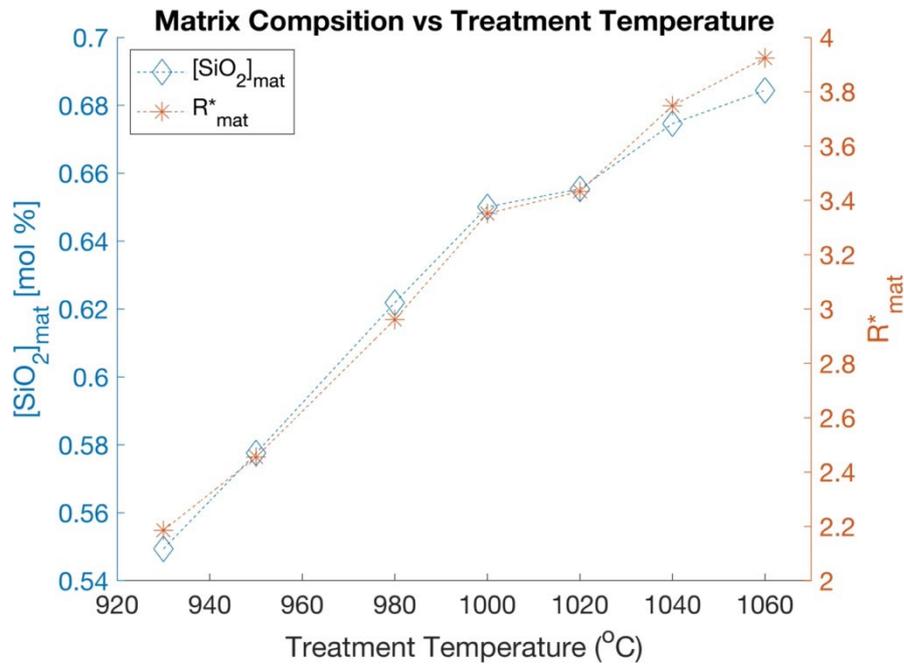

Figure 2: $[SiO_2]$mat and R*mat versus treatment temperature. Both $[SiO_2]$mat and R*mat monotonically increase with treatment temperature

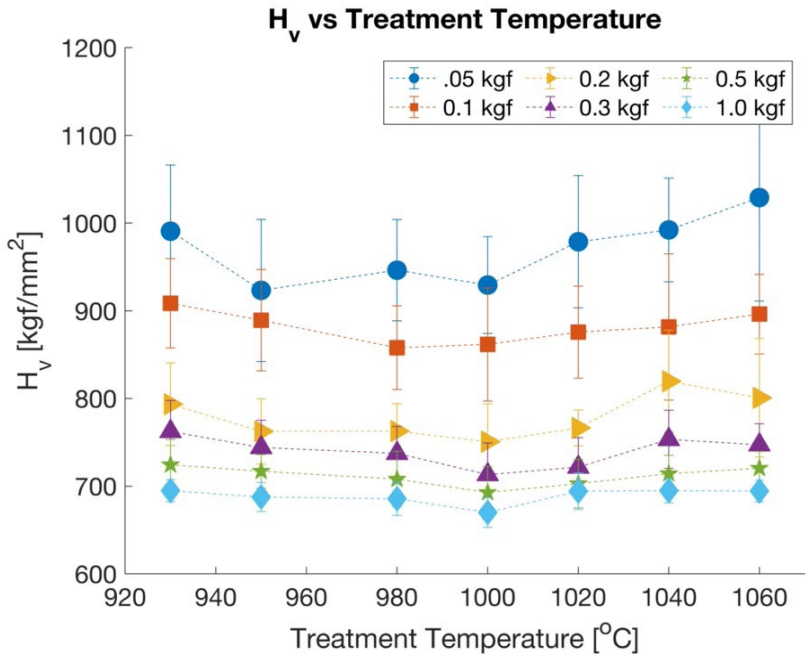

Figure 3: $H_v$ vs treatment temperature. Hardness initially decreases with temperature, then begins to increase beyond 1000°C

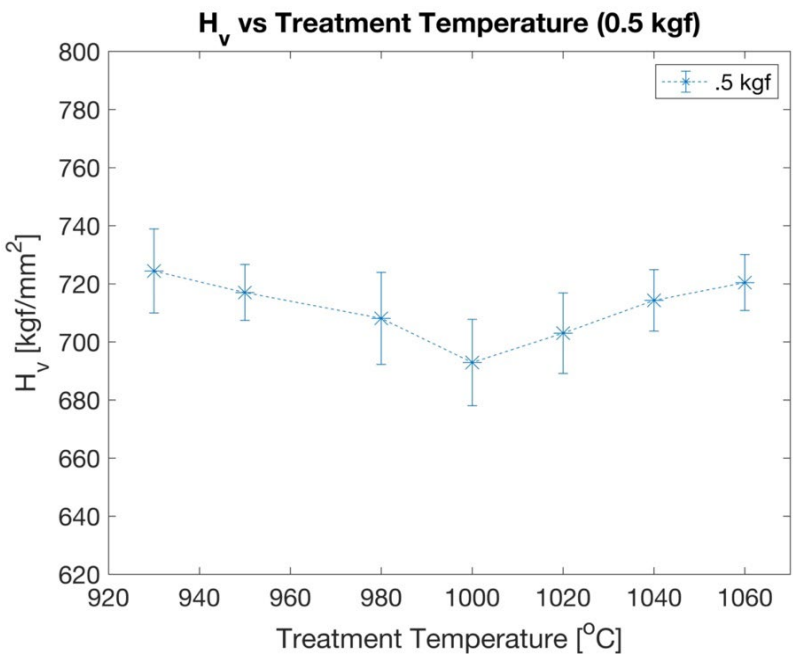

Figure 4: $H_v$ vs treatment temperature at 0.5kgf. The non-monotonic relation between hardness and treatment temperature is more evident when each force is plotted alone

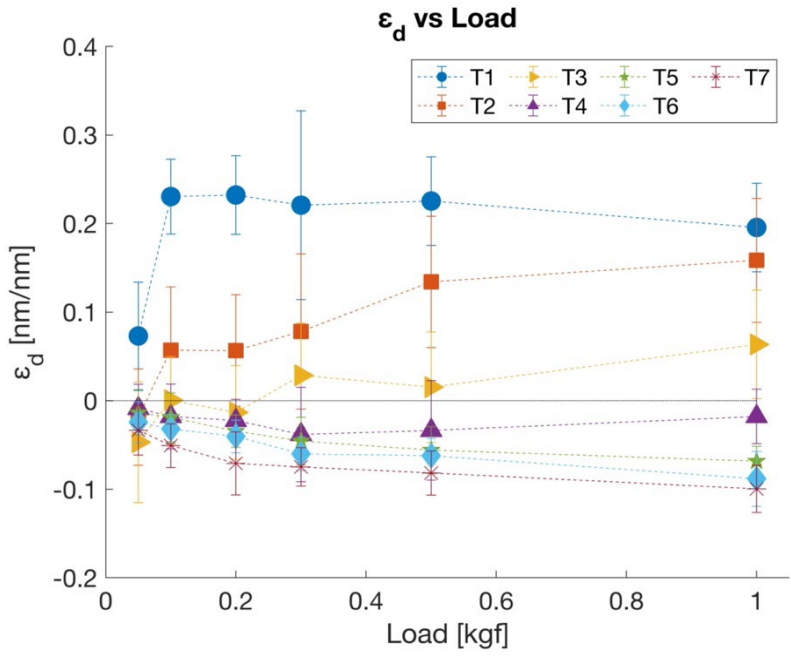

Figure 5: Linear droplet strain as a function of load for all glasses

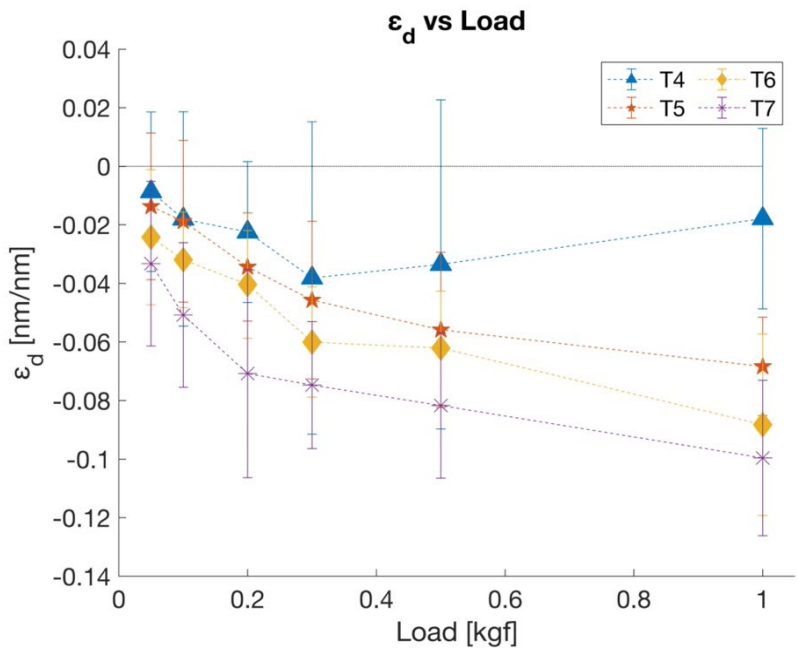

Figure 6: Linear droplet strain for glasses T4, T5, T6, and T7. Droplet size generally decreases with load, except in glass T4

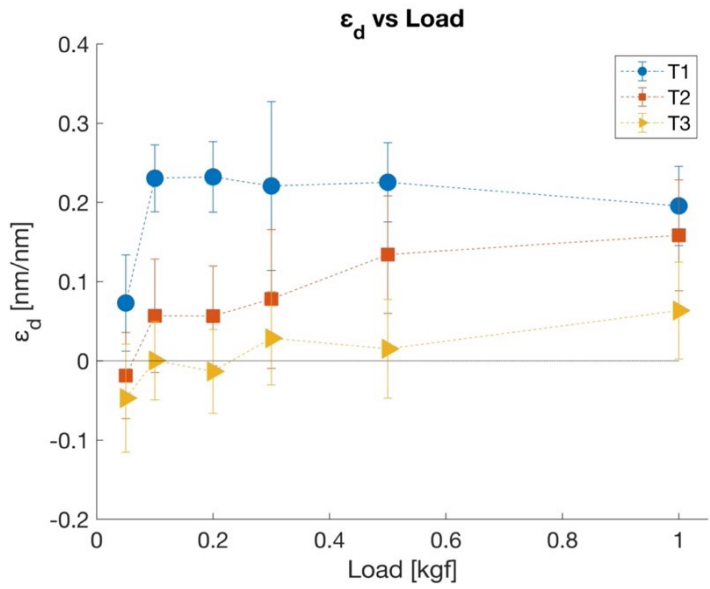

Figure 7: Linear droplet strain as a function of load for glasses T1, T2, and T3

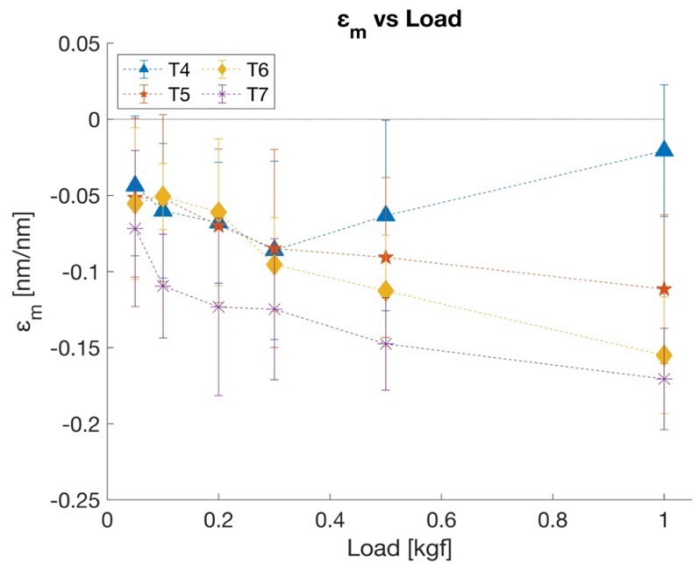

Figure 8: Linear matrix strain for glasses T4, T5, T6, and T7
23

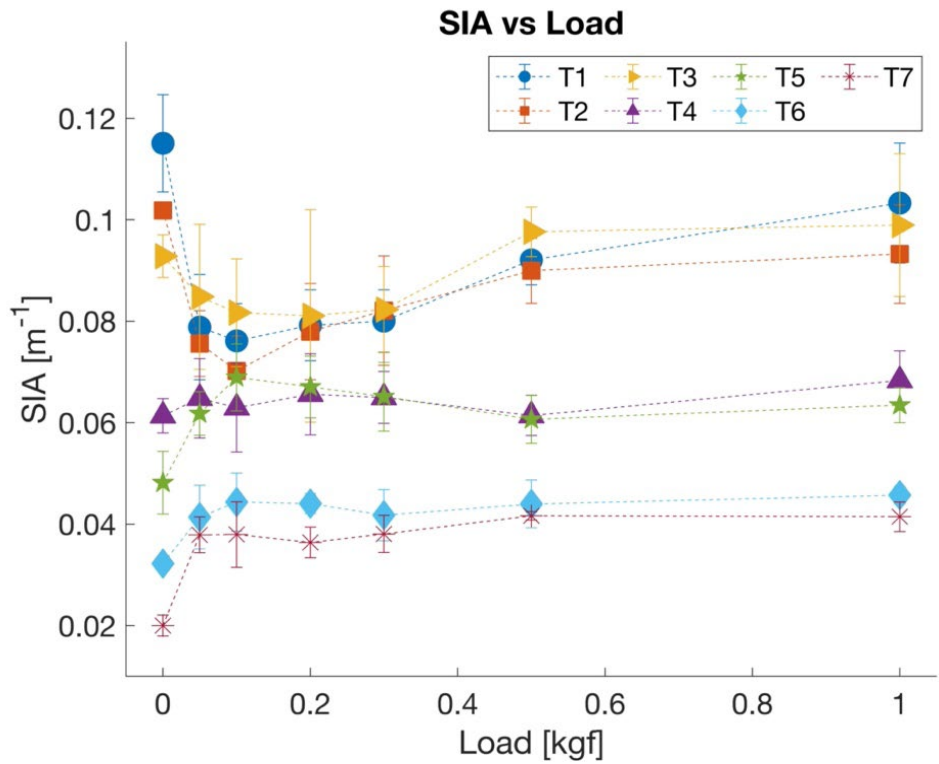

Figure 9: SIA as a function of load for glasses T1-T7

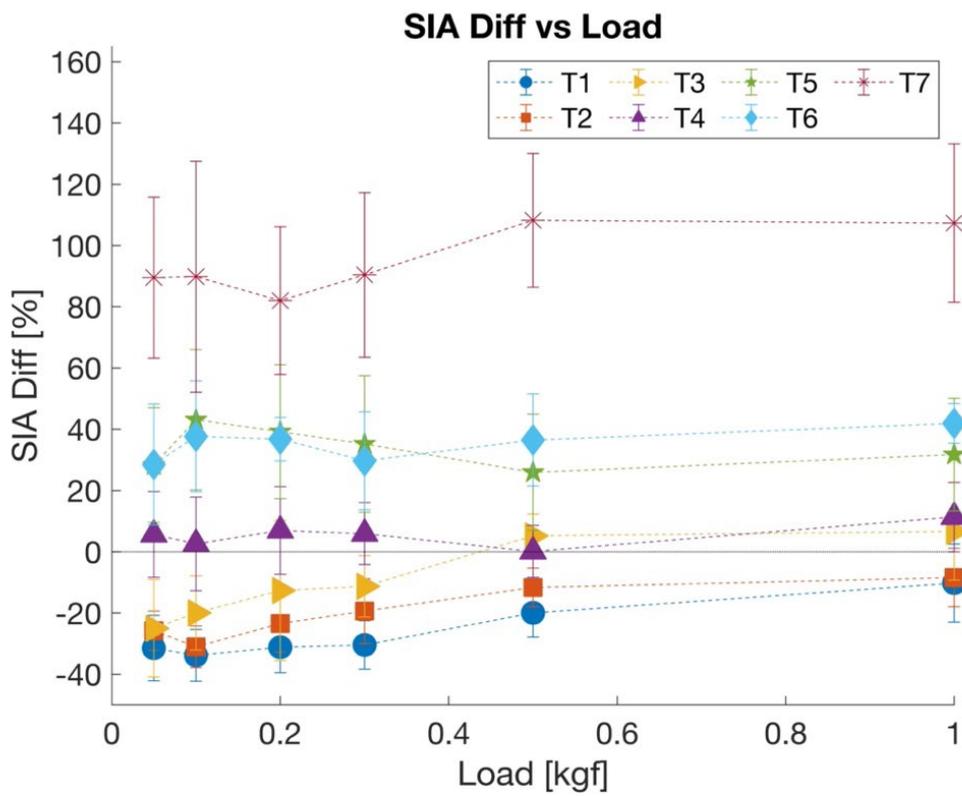

Figure 10: Relative SIA difference for glasses T1-T7. Indentation decreases SIA in glasses T1-T3, increases SIA in T5-T7, and has minimal net effect in glass T4

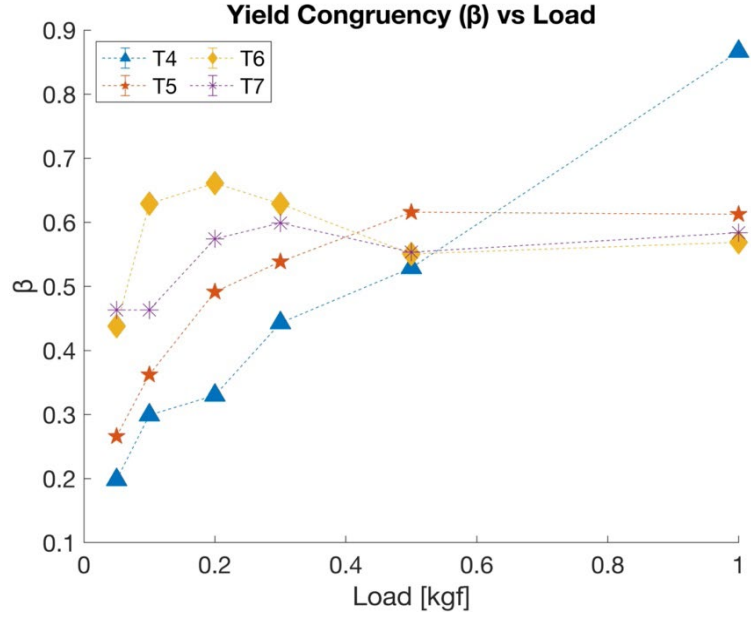

Figure 11: Yield congruency ratio for glasses T4-T7

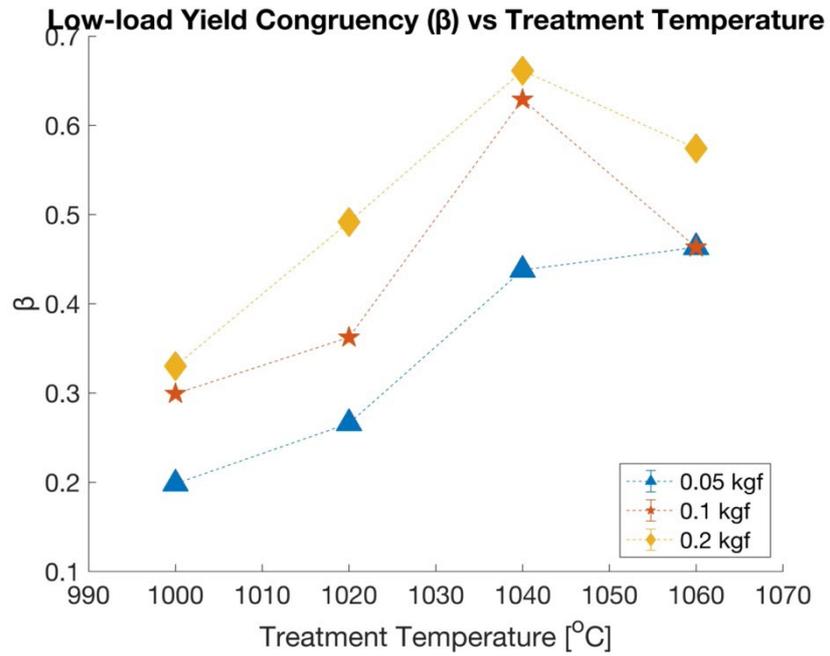

Figure 12: Yield congruency ratio generally increases at "low" loads.

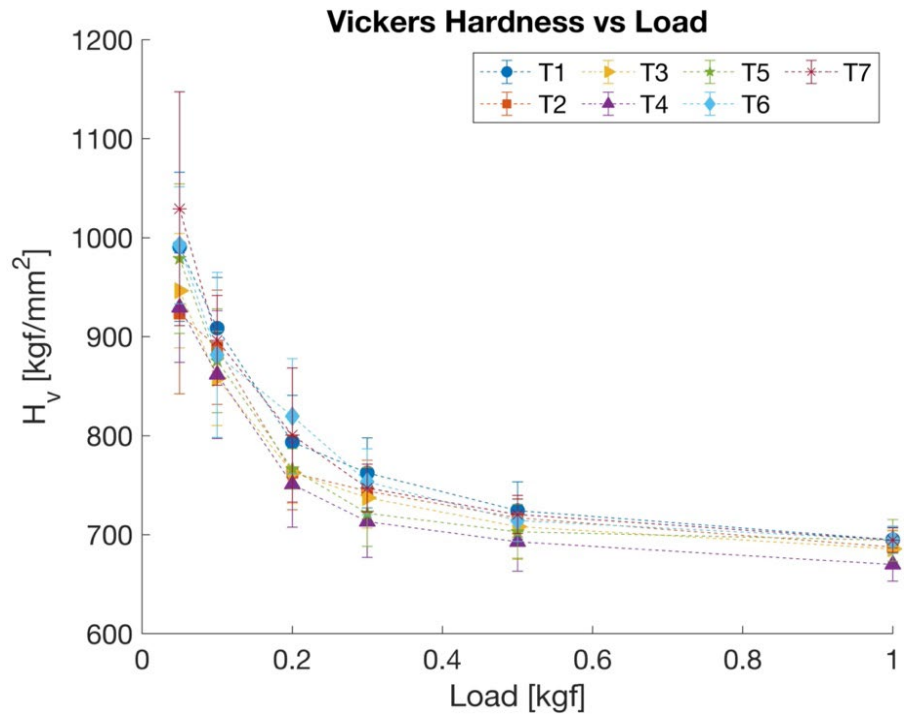

Figure 13: Hardness-Load curves for all T-series glasses

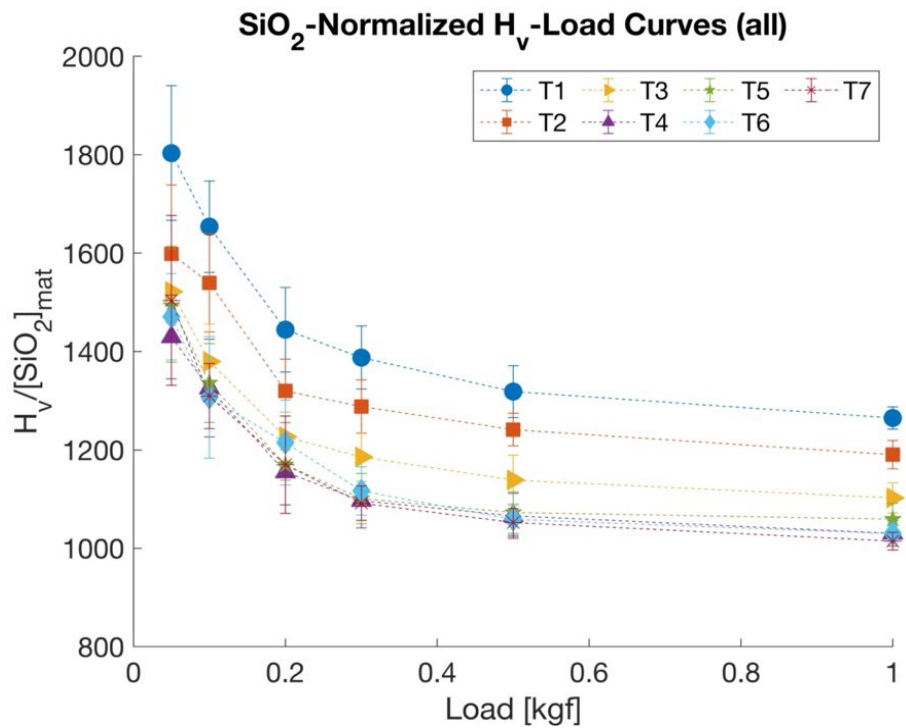

Figure 14: SiO$_2$-normalized hardness-load curves for all T-series glasses. Normalizing hardness by the matrix silica content results in glasses T4-T7 collapsing onto a single hardness-load curve, while glasses T1-T3 diverge

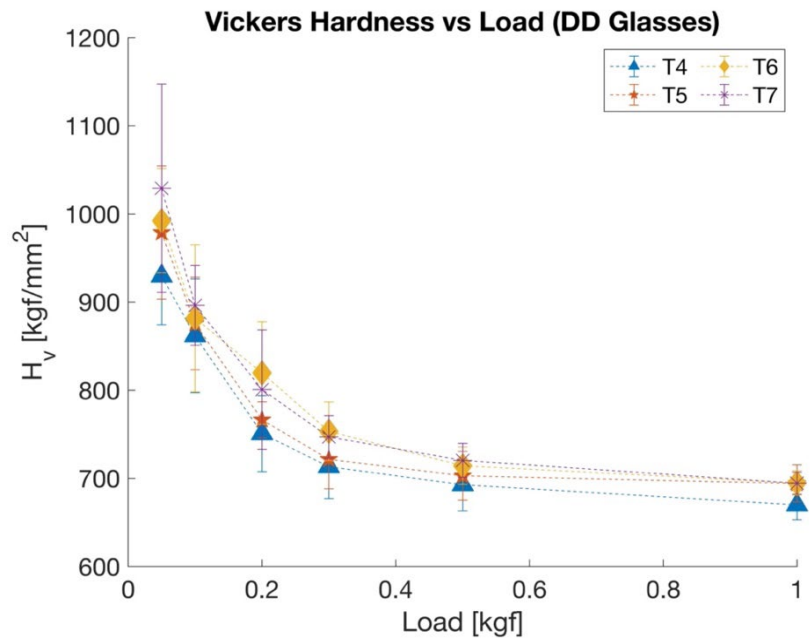

Figure 15: Hardness-load curves for the DD glasses before normalizing to the matrix silica content

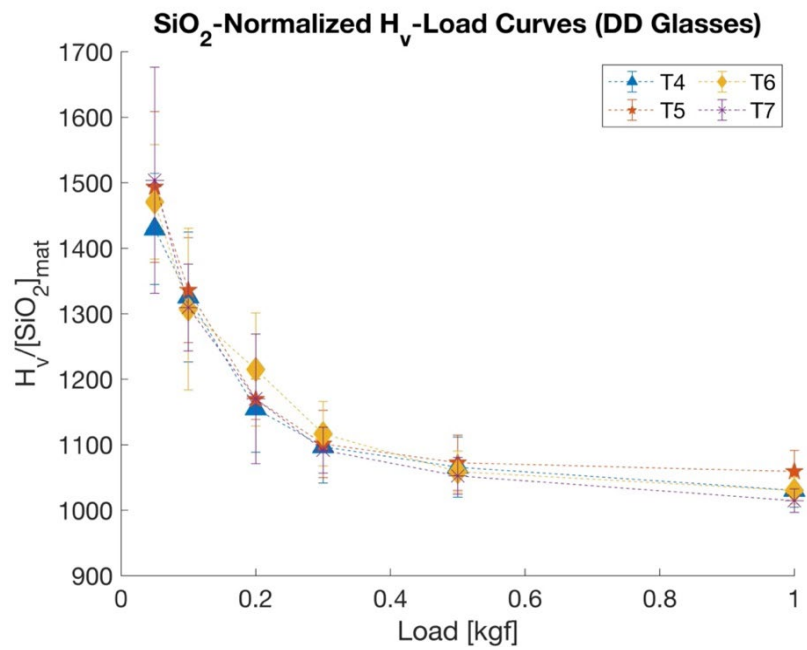

Figure 16: Hardness-load curves for the DD glasses after normalizing hardness by the matrix silica content. Normalizing collapses the hardness-load curves, and indicates matrix silica is responsible for variations in hardness for the DD mechanism

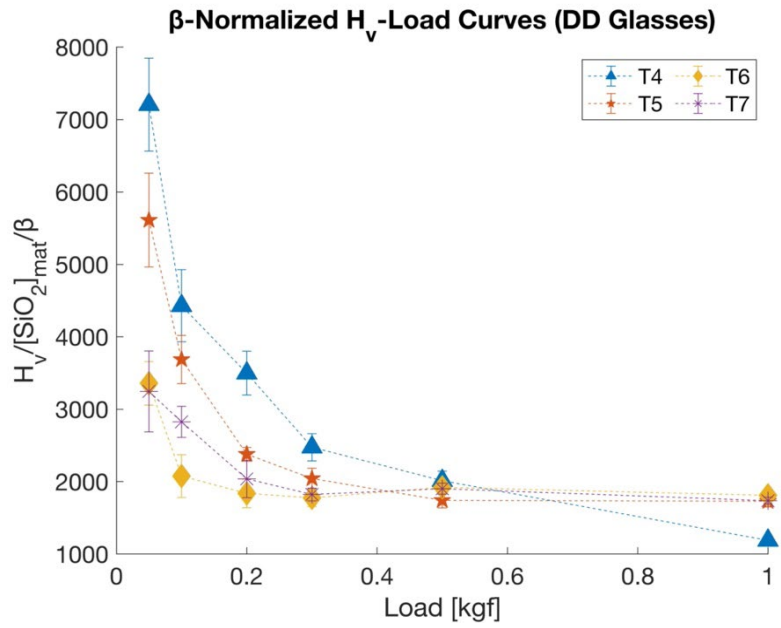

Figure 17: Further normalizing the $H_v$-load curves by beta results in load-independent hardness beyond 0.2kgf for glasses T5, T6, T7. This suggests varying rates of incongruent yielding may be responsible for load-dependent hardness at high loads

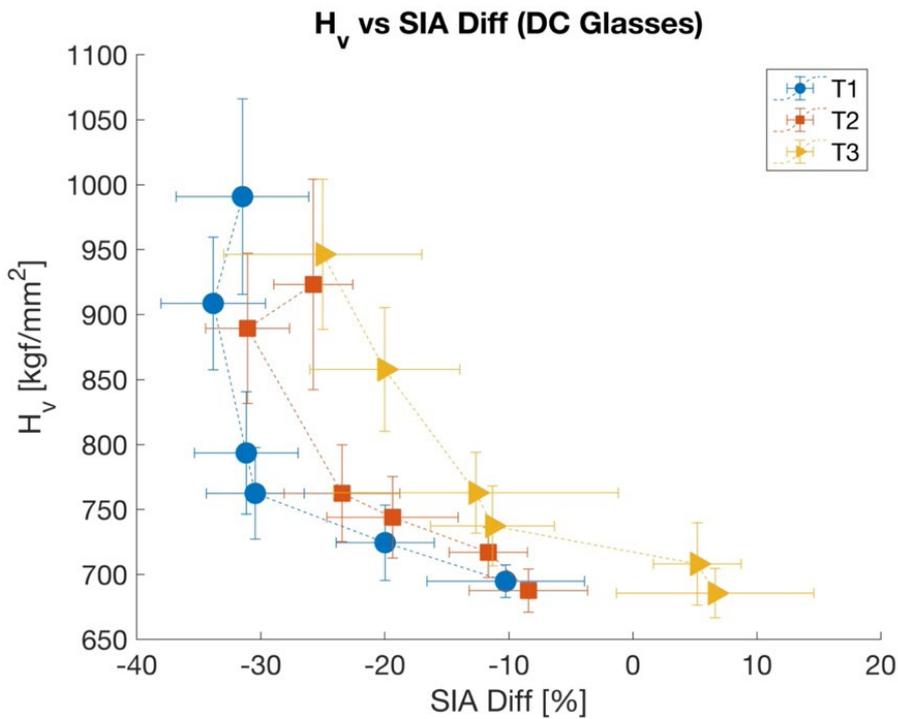

Figure 18: $H_v$ vs relative SIA difference for glasses T1, T2, and T3. Hardness generally increases with greater SIA reduction

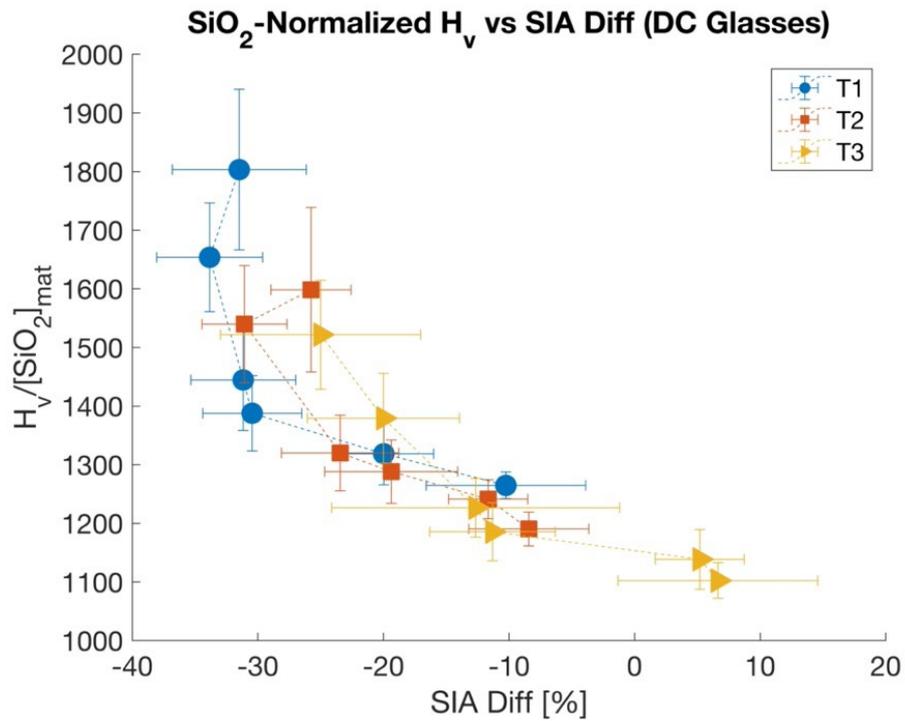

Figure 19: Normalizing $H_v$ by the matrix silica content collapses the individual curves

Table 1: Sample IDs and their corresponding heat treatments

| Sample ID | T1 | T2 | T3 | T4 | T5 | T6 | T7 |
|---|---|---|---|---|---|---|---|
| Treatment Temperature (°C) | 930 | 960 | 980 | 1000 | 1020 | 1040 | 1060 |
| Soak Time (minutes) | 30 | 30 | 30 | 30 | 30 | 30 | 30 |

Table 2: Estimated matrix composition

| Temperature (°C) | $SiO_2$ (mol%) | $CaO$ (mol%) | $Al_2O_3$ (mol%) |
|---|---|---|---|
| 930 | 0.5494 | 0.3509 | 0.0995 |
| 950 | 0.5776 | 0.3288 | 0.0934 |
| 980 | 0.6219 | 0.294 | 0.084 |
| 1000 | 0.6500 | 0.2719 | 0.0779 |
| 1020 | 0.6554 | 0.2677 | 0.0768 |
| 1040 | 0.6746 | 0.2526 | 0.0727 |
| 1060 | 0.6844 | 0.245 | 0.0706 |